\documentclass{mn2e}
\usepackage{graphicx}
\usepackage{amsbsy}
\usepackage{amsmath}
\usepackage{epsf}

\newcommand{\apj}{ApJ}
\newcommand{\apjl}{ApJL}
\newcommand{\mnras}{MNRAS}
\newcommand{\apjs}{ApJS}
\newcommand{\pasp}{PASP}
\newcommand{\aap}{A\&A}

\newcommand{\beq}{\begin{equation}}
\newcommand{\eeq}{\end{equation}}

\DeclareMathAlphabet{\mathsfsl}{OT1}{cmss}{bx}{sl}
\SetMathAlphabet{\mathsfsl}{bold}{OT1}{cmss}{bx}{sl}

\newcommand\phs{\phantom{$-$}}%
\newcommand\phn{\phantom{0}}%
\newcommand\arcdeg{\mbox{$^\circ$}}%
\newcommand{\hour}{\hbox{$^{\rm h}$}}
\newcommand{\minute}{\hbox{$^{\rm m}$}}

\newcounter{fn}

\usepackage{epstopdf}
\begin{document}

\title[Shapes and B-Field Orientations in Molecular Clouds]
{Statistical Assessment of Shapes and Magnetic Field
  Orientations in Molecular Clouds through Polarization Observations}
%

\author[K. Tassis et al.]
  {K.~Tassis$^{1}$, C.~D.~Dowell$^1$, R.~H.~Hildebrand$^{2,3}$,
    L.~Kirby$^2$, and J.~E.~Vaillancourt$^4$ \\
    $^1$Jet Propulsion Laboratory, California Institute of Technology, Pasadena, CA 91109\\
    $^2$Enrico Fermi Institute and Department of Astronomy and
    Astrophysics, University of Chicago, 5640 South Ellis Avenue,
    Chicago, IL 60637 \\
    $^3$Department of Physics, University of
    Chicago, 5720 S. Ellis Ave. Chicago, IL 60637\\
    $^4$Division of Physics, Mathematics, \& Astronomy, California
    Institute of Technology, Pasadena, CA 91125}

\maketitle 

\label{firstpage}
\begin{abstract}
We present a novel statistical analysis aimed at deriving the intrinsic
shapes and magnetic field orientations of molecular clouds using dust
emission and polarization observations by the Hertz polarimeter. Our
observables are the aspect ratio of the projected plane-of-the-sky
cloud image, and the angle between the mean direction of the
plane-of-the-sky component of the magnetic field and the short axis of
the cloud image.  To overcome projection effects due to the unknown
orientation of the line-of-sight, we combine observations from 24
clouds, assuming that line-of-sight orientations are random and all are
equally probable.  Through a weighted least-squares analysis, we find
that the best-fit intrinsic cloud shape describing our sample is an oblate
disk with only small degrees of triaxiality. The best-fit intrinsic
magnetic field orientation is close to the direction of the shortest
cloud axis, with small ($\sim 24^\circ$) deviations toward the
long/middle cloud axes. However, due to the small number of observed
clouds, the power of our analysis to reject alternative configurations
is limited.
\end{abstract}

\section{Introduction}

Far-infrared and sub-millimeter emission from molecular clouds appears
polarized, presumably as a result of the alignment of elongated dust
grains with the cloud magnetic field (see, e.g., Curran \& Chrysostomou 2007; 
Dotson et al.\ 2000, 2009; Hildebrand et
al.\ 2000;   Hoang \& Lazarian 2008; Lazarian 2003, 2007). Measurable degrees of polarization, at the 
few percent level, are typical in many interstellar clouds and cloud
cores (Dotson et al.\ 2000, 2009). Additionally, theoretical
advancements in understanding grain alignment 
(e.g., Bethell et al.\ 2007) indicate
that even in relatively dense clouds, dust polarization traces 
the underlying magnetic field structure.  The ubiquity of
magnetic fields and dust in the interstellar medium makes polarization
observations a powerful tool for distinguishing between theories of
molecular cloud formation, support, and evolution.

From a theoretical point of view, the orientation of the mean magnetic
field in molecular clouds is closely tied to the dynamical importance
of magnetic forces compared to gravity, random motions (turbulence),
and thermal pressure. If magnetic fields are dynamically important, and
are responsible for a significant fraction of the support of molecular
clouds against gravity, then the mean magnetic field is preferentially
oriented parallel to the shortest axis of the molecular cloud (e.g.,
Mouschovias 1978).  This is a result of an increased support against
gravity in the direction perpendicular to the magnetic field compared
to the direction parallel to it. The cloud contracts
more in the direction parallel to the magnetic field than in the
direction perpendicular to the field.
An additional result of magnetic support is that the intrinsic shapes
of molecular clouds in this case resemble mostly oblate (although not
necessarily axisymmetric), flattened ellipsoids (one axis appreciably
smaller than the other two).

If magnetic fields are dynamically unimportant compared to turbulence,
then turbulent motions dominate the internal dynamics of clouds, and
the magnetic fields are dragged around by turbulent eddies (e.g.,
Ballesteros-Paredes, V\'{a}zquez-Semadeni \& Scalo 1999). In this
case, the mean magnetic field has a random orientation with respect to
the molecular cloud principal axes. The shape distribution for
overdensities formed in a turbulent field is also random (Gammie et
al.\ 2003). Molecular clouds forming out of magnetohydrodynamic
turbulence in the weak-field regime therefore would be expected to
have random shapes and random magnetic field orientations.

A third possibility is that magnetic fields have a helical
configuration and thread prolate (filamentary) molecular clouds (Fiege
\& Pudritz 2000a).  The most common outcome of such configurations is
polarization patterns that may contain $90^\circ$ flips of the
polarization vector (Fiege \& Pudritz 2000b). Additional ideas for
molecular cloud shapes come from non-magnetic calculations; for
example, finite, self-gravitating gaseous sheets have been shown to
collapse to filamentary structures with mass concentrations close to
the edges or ends of the filaments (e.g., Burkert \& Hartmann 2004;
Hartmann \& Burkert 2007).

If it were possible to determine the mean orientation of the ordered
magnetic field in molecular clouds, then important constraints could
be placed on theories of molecular cloud formation and
support. However, such a task is not straightforward, due to various
projection effects which prohibit us from knowing either the true
orientation of the magnetic field or the principal axes of the cloud
ellipsoid. Polarization measurements can only determine the
orientation of the magnetic field on the plane-of-the-sky (POS), and,
similarly, only the shape of the POS projection of the cloud can be
measured. As a result, the magnetic field may be oriented, for
example, along the shortest cloud axis, but its projection on the sky
may appear closest to the major axis of the projected cloud
ellipse. These difficulties were explicitly demonstrated by Basu
(2000).

An effective way to overcome these difficulties and use polarization
measurements to constrain theoretical models for the orientation of
the mean magnetic field in molecular clouds is through a statistical
treatment. Assuming that the orientation of the clouds themselves with
respect to our LOS is random, we can assess the likelihood
of different underlying distributions of molecular cloud shapes, and
of orientations of the mean magnetic field in molecular clouds. In
this work, we present a new method for such a treatment. We study a
sample of 24 molecular clouds with measured apparent elongations and
field orientations. By applying a weighted least-squares analysis, we
derive the most probable intrinsic shapes and most probable magnetic
field orientations for the parent population that the 24 clouds
sample.

This paper is organized as follows. The observations and analysis used
to derive elongations and apparent field orientations in the sample
clouds are described in \S\ref{obs}. The formalism used in our
analysis, including a short discussion of projection effects and a
description of our statistical analysis, are discussed in \S\ref{form}
(a detailed mathematical treatment of projection effects and the
statistical analysis are given in Appendices \ref{approj} and
\ref{apstat} respectively). Our results are presented in \S\ref{res},
and discussed in \S\ref{disc}.

\section{Observations} \label{obs}

The Hertz polarimeter (Schleuning et al.\ 1997; Dowell et al.\ 1998)
deployed at the Caltech Submillimeter Observatory has been used to map
the flux density and polarization of molecular clouds on scales of
several arcminutes at a wavelength of $350\,\mu$m.  In a complete
sample of 56 Galactic clouds, 32 contain 2 or more polarization measurements
satisfying the criterion $P \geq 3\sigma_p$ or better, where $P$ is
the measured polarization amplitude and $\sigma_p$ is its measurement
uncertainty.\footnote{In counting 32 clouds, we have counted
  separately two components of OMC-3 with distinct cores separated by
  more than $2\arcmin$, with no Hertz measurement of flux in between.}  The clouds
are listed in Table \ref{datatable}; object positions and complete
maps of flux density and polarization can be found in Dotson et
al.~(2009).

\subsection{Polarization} \label{sec-poldata}

\begin{figure}
\includegraphics[width=0.9\columnwidth]{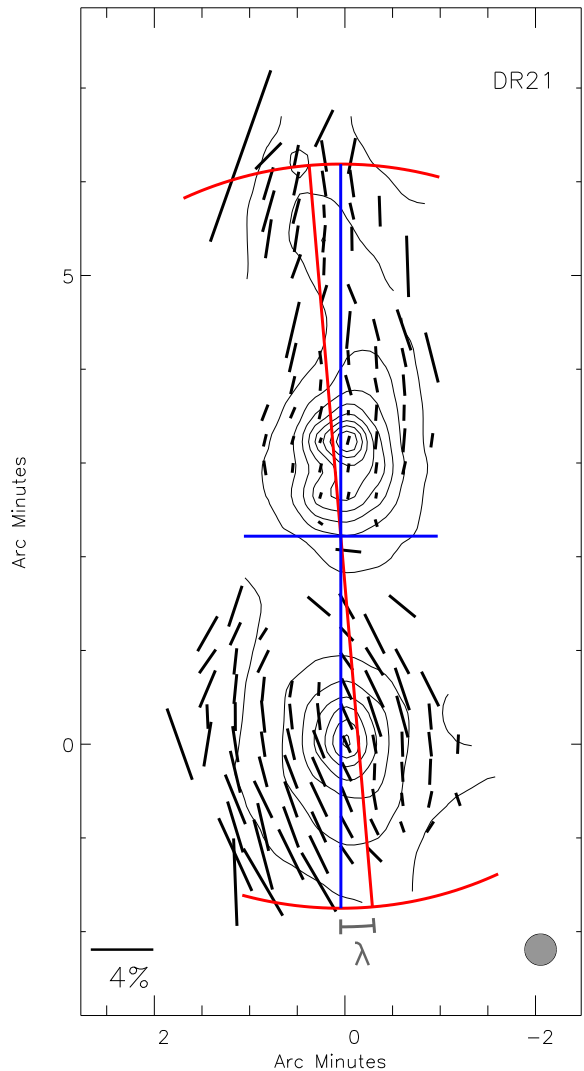}
\caption{Mean polarization and cloud shape in DR21. Polarization
  measurements (small black line-segments) are plotted with their lengths
  proportional to $P$ (scale at lower-left); only $P \geq 3\sigma_p$
  measurements are shown. The mean polarization direction ($E$-vector) and
  its dispersion are shown as the red line and the arcs at its end.
  The cloud principal axes are drawn as blue lines; their relative
 lengths are given by the aspect ratio $q = 0.26$. 
The lines corresponding to the mean polarization and cloud widths  
intersect at the cloud center ($x_0$,$y_0$)
  and are drawn with arbitrary absolute lengths.  The gray arc labeled
  ``$\lambda$'' denotes the difference between the mean polarization
  angle and the cloud position angle.  Coordinate offsets from
  $20\hour39\minute1\fs1$, $ 42\arcdeg19\arcmin31\arcsec$ (J2000).
  Contours are 10, 20, ..., 90\% of the peak flux density of
  820\,Jy/beam (at ($\Delta\alpha, \Delta\delta$) = (0\arcmin,
  3\farcm3)).  The shaded circle at lower-right is the $20\arcsec$
  FWHM Hertz beam.}
\label{dr21}
\end{figure}

\begin{figure}
\includegraphics[width=0.9\columnwidth]{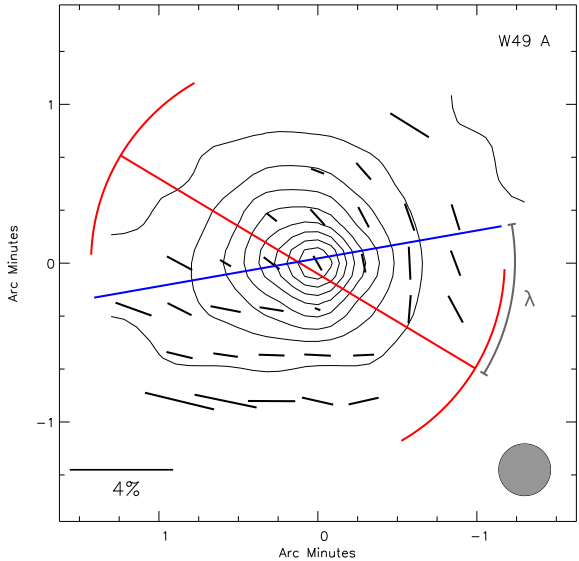}
\caption{Same as Figure \ref{dr21} but for the cloud W49\,A. As this
  cloud is nearly circular (aspect ratio $q=0.87$) we plot only the
  long cloud principal axis for clarity. Coordinate offsets are from
  $19\hour10\minute13\fs6$, $ 9\arcdeg 6\arcmin17\arcsec$ (J2000).
  Contours are 10, 20, ..., 90\% of the peak flux density of
  730\,Jy/beam.}
\label{g34}
\end{figure}

Within each cloud, we limit the data points used to calculate mean
polarizations to those satisfying the condition $P \geq 3\sigma_p$.
The best estimate of the POS-projected magnetic field direction is
that of the measured polarization vector (rotated by $90^\circ$).  The
amplitude of the polarization $P$ yields some information about the
inclination of the field to the line-of-sight (LOS\@).  However, $P$
is also dependent on a number of other factors which are independent
of the field, including the dust grain shape and alignment efficiency, and the magnetic field structure along the LOS and within the
telescope beam (e.g., Draine \& Lee 1985).
Therefore, our analysis of the magnetic field will involve only the
polarization position angle.

The mean polarization position angle $\langle \chi \rangle$ is defined
by averaging Stokes parameters of unit magnitude such that
\begin{equation}
\langle \chi \rangle \equiv
\frac{1}{2}\tan^{-1}\frac{\langle \sin 2\chi\rangle}{\langle \cos 2\chi\rangle},
\label{eq-avgchi}
\end{equation}
 where
\begin{equation}
\langle \sin 2\chi\rangle = \frac{1}{N} \sum_{i=1}^N \sin 2\chi_i,
\end{equation}
\begin{equation}
\langle \cos 2\chi\rangle = \frac{1}{N} \sum_{i=1}^N \cos 2\chi_i,
\label{eq-avgcos}
\end{equation}
and the $\chi_i$'s are the measured polarization angles at each
position $i$ in the cloud.  The sums are over all $N$ points  in the
map where the measured polarization $P \geq 3\sigma_p$.  

We calculate the standard deviation of the mean $\langle \chi
\rangle$ under the assumption of zero intrinsic dispersion (i.e.,
under the assumption that all of the dispersion in the observed
$\chi_i$ is due to observational errors) as
\begin{equation}
\sigma_{\langle \chi \rangle} = \frac{1}{N} \left[ \sum_{i=1}^N \sigma_i^2 \right] ^{1/2}
\end{equation}
where the $\sigma_i$ are the uncertainties on the measurements of
$\chi_i$.  The measurement uncertainty is limited to Hertz's
systematic uncertainties such that $\sigma_{\langle \chi \rangle} \geq
2^{\circ}$ (Dowell et al.\ 1998).  In our case, the dispersion in the
observed values of $\chi_i$ in individual clouds is substantial, so we
also calcluate the dispersion about the mean, $\sigma_\chi$, given by
the standard deviation of the measurements
\begin{equation}
\sigma_\chi = \left[ \frac{1}{N-1} \sum_{i=1}^N \left(\chi_i - \langle \chi \rangle\right)^2 \right]^{1/2}.
\label{eq-disp}
\end{equation}
In using this equation we have accounted for the 180-degree periodicity in the 
polarization angles.

Each cloud's mean polarization amplitude is given by
\begin{equation}
\langle P \rangle \equiv \frac{\sqrt{\langle Q \rangle^2 + \langle U \rangle^2}}{\langle I \rangle}
\label{eq-avgpol}
\end{equation}
where $I$, $Q$, and $U$ are the measured Stokes parameters. The mean
polarizations, angles, and uncertainties are all reported in Table
\ref{datatable}\@.  The $\langle P \rangle$'s reported in Table
\ref{datatable} have been de-biased using the statistical measurement
uncertainty on the mean (e.g., Vaillancourt 2006). However, the
measurement uncertainties in the table are not those used to de-bias,
but are limited to Hertz's systematic uncertainties such that
$\sigma_{\langle p \rangle} \geq 0.2\%$ (Dowell et al.\ 1998).

\subsection{Cloud Shapes} \label{sec-shapedata}

In the case of a symmetric ellipsoidal cloud, the first- and
second-moments of the flux density describe the cloud's center and
width.  The location of the cloud's center $(x_0,y_0)$ is given by the
first-moments:
\begin{eqnarray}
x_0 & = & \frac{\sum_i x_i\; F(x_i,y_i)}{\sum_i F(x_i,y_i)}, \label {eq-comx} \\
y_0 & = & \frac{\sum_i y_i\; F(x_i,y_i)}{\sum_i F(x_i,y_i)}.
\label{eq-comy}
\end{eqnarray}
where the sum is over all points $i$ in the flux density map.  No cuts
are applied to the flux density data beyond those already made in the
Dotson et al. (2009) archive. (The $P\geq3\sigma_p$ criterion is not
applied.)

The second moments compose the 3 elements of a symmetric $2\times2$ matrix
\begin{equation}
\mathbf{I} = \left(\begin{array}{cc} I_{xx} & I_{xy} \\ I_{xy} & I_{yy} \end{array} \right)
\end{equation}
with elements
\begin{eqnarray}
I_{xx} & = & \phantom{-} \sum_i (y_i-y_0)^2\; F(x_i,y_i), \\
I_{xy} & = & - \sum_i (x_i-x_0)(y_i-y_0)\; F(x_i,y_i), \\
I_{yy} & = &  \phantom{-} \sum_i (x_i-x_0)^2\; F(x_i,y_i).
\label{eq-momyy}
\end{eqnarray}
The principal moments of the cloud are simply the eigenvalues, $I_1$
and $I_2$, of $\mathbf{I}$. Defining $I_1 \geq I_2$, the aspect ratio
is given by $q= \sqrt{I_2/I_1}$ and the position angle of the long
cloud axis is determined from the eigenvector
corresponding to the eigenvalue $I_2$.

The first and second moments are well-defined and express the
distribution of the flux density in the cloud even when the shape is
irregular and cannot be well approximated by an ellipse.
Therefore, we use these moments to define the aspect
ratio and position angle of the principal axes for all clouds in our sample, with the caveat
that we have ignored any higher-order moments which more precisely
characterize the cloud shape.
We note that some bias can be introduced through 
the finite size of each map. 

\begin{figure}
\includegraphics[width=0.9\columnwidth]{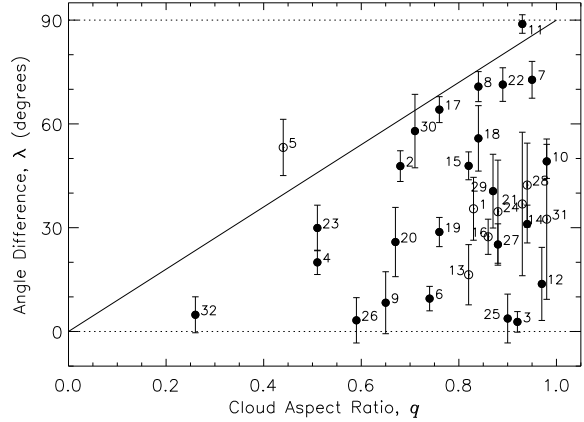}
\caption{Elongations and angles between the magnetic field and the
  minor cloud axis for the 32 clouds in our sample. Note that the
  elongation $q$ is defined so that a very elongated cloud corresponds
  to $q\rightarrow 0$ and a circular cloud to $q=1$. For
  angles, $\lambda = 0$ corresponds to a projected magnetic field
  aligned with the minor axis of the projected cloud.  Each error bar
  represents the quadrature sum of the statistical uncertainty of the
  cloud's mean angle uncertainty ($\sigma_{\langle\chi\rangle}$), the
  systematic angle uncertainty, and the systematic polarization
  uncertainty (see text for precise definitions of these terms).
Data shown as open circles are not included in further analysis due to
their large dispersion in $\lambda$ (all data are given in Table
\ref{datatable}). Dotted horizontal lines are drawn at the boundaries
of the allowed range $0^\circ < \lambda  < 90^\circ$. The diagonal $\lambda/90^\circ = q$ is shown as a solid line.}
\label{FIGURE1}
\end{figure}

Examples of the mean polarization and cloud shape calculations are
shown in Figure~\ref{dr21} for the most elongated cloud, and
Figure~\ref{g34} for a nearly circular cloud.  The measured cloud
parameters are given in Table~\ref{datatable}. The apparent cloud
aspect ratio $q$ and the absolute value of the angle between mean
field and short apparent axis, $\lambda$, are plotted in
Fig.~\ref{FIGURE1}. Data shown as open circles in Fig.\ \ref{FIGURE1}
and denoted by $h$ in Table \ref{datatable} are not included in
further analysis due to their large ($> 30^\circ$) dispersion in
$\lambda$.

Error bars on $\lambda$ correspond to the
quadrature sum of $\sigma_{\langle\chi\rangle}$, the systematic angle
uncertainty (2 degrees; Dowell et al.\ 1998), and the systematic
polarization uncertainty ($\Delta\chi_{\mbox{sys}}$).  We estimate the
last quantity using the relation 
\begin{equation}
\Delta\chi_{sys} = \frac{90\arcdeg}{\pi} \frac{0.2\%}{\langle P\rangle},
\label{eq-angsys}
\end{equation} 
where 0.2\% is Hertz's systematic polarization uncertainty.
We note that equation (\ref{eq-angsys}) assumes measurements of $P$
and $\chi$ follow normal distributions (equivalent to the assumption
$P \gg \sigma_p$).  While this is clearly not true for all
measurements in Table \ref{datatable}, this fact has minimal effect on
the apparent size of the error bars and has no effect on our
subsequent analysis (as the uncertainties are not
used).\footnote{Further study of polarization angle uncertainties at
low $P/\sigma_p$ can be found in Naghizadeh-Khouei \& Clarke (1993)}

Furthermore, these angle uncertainty estimates do not include any
uncertainties in the estimate of the cloud shape, but include only the
uncertainties on the measurement of the polarization angle.  A visual
comparison of Figures \ref{dr21} and \ref{g34} should make it apparent
that the uncertainty in the cloud orientation is smallest when $q=0$
and must increase as the cloud becomes more circular ($q=1$). An
estimate of this uncertainty for all clouds in our study is beyond the
scope of this paper.  Neglecting these uncertainties has no effect on
our conclusions as this information is not used in our subsequent data
analysis.

 Data shown as open circles in
Fig.\ \ref{FIGURE1} and denoted by $h$ in Table \ref{datatable} are
not included in further analysis due to their large ($> 30^\circ$)
dispersion in $\lambda$.

%
%
%
%

\begin{table*}
  \begin{minipage}{7.0in}
    \caption{Mean Polarization and Cloud Shape Parameters}
    \label{datatable}
    \begin{tabular}{clcccccccccc}
      \hline
      & & \multicolumn{6}{c}{Mean Polarization Parameters ($E$-vector)%
      \footnote{Mean polarization parameters defined in equations (\ref{eq-avgchi}) -- (\ref{eq-avgpol}). Means are calculated using only the $P \geq 3\sigma_p$ vectors shown in ``No.\ vectors'' column. Angle $\chi$ is measured east of north.}%
      } & & \multicolumn{2}{c}{Cloud Shape}\\
      \cline{3-8} \cline{10-11} \\
      Object & Object\footnote{Object names follow the convention in Dotson et al. (2009); list is ordered with approximately increasing Right Ascension.}%
      & No. & $\langle P \rangle$ & $\sigma_{\langle p \rangle}$ & 
      $\langle \chi \rangle$%
      &$\sigma_{\langle \chi \rangle}$ & 
      $\sigma_\chi$\footnote{Dispersion about mean polarization angle given by equation (\ref{eq-disp}).}
      & & Aspect Ratio & Angle\footnote{Direction of long cloud axis, measured east of north} &
      $\vert \lambda \vert$\footnote{Angle between mean polarization angle and long cloud axis.  This is equivalent to the angle between the inferred magnetic field direction and the short cloud axis.} \\
      No.   & Name%
      & vectors & (\%) & (\%) & (degrees) &    (degrees) &  (degrees) & & & (degrees) & (degrees) \\      \hline
\setcounter{fn}{8}
   \phn1 &              W3$^\alph{fn}$                           & \phn51 & 0.6  &  0.2  & \phs 65      &   \phn2   &  32   &  & 0.83 & $-$80     &  35   \\
   \phn2 &    NGC 1333                                & \phn18 & 1.6  &  0.2  & \phs 88      &   \phn2   &  19   &  & 0.68 & $-$45     &  48   \\
   \phn3 & IRAS 05327-0457                            & \phn18 & 3.9  &  0.4  & $-$40        &   \phn2   &  17   &  & 0.92 & $-$37     & \phn3 \\
   \phn4 &           OMC-1                            & 437    & 2.0  &  0.2  & \phs 25      &   \phn2   &  26   &  & 0.51 & \phs\phn5 &  20   \\
   \phn5 &           OMC-2$^\alph{fn}$                            & \phn26 & 0.7  &  0.2  & $-$39        &   \phn4   &  33   &  & 0.44 &  \phs 14  &  53   \\
   \phn6 &      OMC-3 MMS\,6%
     \footnote{Part of OMC-3 in Dotson et al. (2009)} & \phn43 & 2.1  &  0.2  & $-$42        &   \phn2   &  12   &  & 0.74 & $-$51     & \phn9 \\
\setcounter{fn}{6}
   \phn7 &      OMC-3 MMS\,9$^\alph{fn}$               & \phn21    & 1.2  &  0.2  & $-$62        &   \phn2   &  20   &  & 0.95 &  \phs 11  &  73   \\
   \phn8 &           OMC-4                            & \phn16    & 1.7  &  0.2  & $-$14        &   \phn2   &  13   &  & 0.84 &  \phs 57  &  71   \\
   \phn9 &    NGC 2024                                & \phn54    & 0.7  &  0.2  & $-$23        &   \phn2   &  24   &  & 0.65 & $-$15     & \phn8 \\
      10 &  NCG 2068 LBS17                            & \phn\phn3 & 8.3  &  1.5  & \phs 81      &   \phn5   &  16   &  & 0.98 & $-$50     &  49   \\
      11 &  NCG 2068 LBS10                            & \phn43    & 3.8  &  0.2  & $-$51        &   \phn2   &  20   &  & 0.93 & \phs38    &  89   \\
      12 &        NGC 2071                            & \phn\phn6 & 0.6  &  0.2  & $-$27        &   \phn4   &  25   &  & 0.97 & $-$13     &  14   \\
\setcounter{fn}{8}
      13 &          Mon R2$^\alph{fn}$                            & \phn49    & 0.7  &  0.2  & \phs 31      &   \phn2   &  32   &  & 0.82 & \phs47    &  16   \\
      14 &           GGD12                            & \phn17    & 1.2  &  0.2  & \phs 89      &   \phn2   &  16   &  & 0.94 & $-$60     &  31   \\
      15 &            S269                            & \phn\phn7 & 2.9  &  0.3  & \phs 30      &   \phn3   &  13   &  & 0.82 & $-$18     &  48   \\
      16 &        AFGL 961$^\alph{fn}$                            & \phn\phn5 & 1.9  &  0.7  & $-$10        &   \phn4   &  44   &  & 0.86 & $-$37     &  27   \\
      17 & Mon OB1 12\footnote{a.k.a.\ IRAS 06382+0939} & \phn26  & 2.0  &  0.2  & $-$37        &   \phn2   &  17   &  & 0.76 & \phs27    &  64   \\
      18 &        NGC 2264                             & \phn18   & 0.6  &  0.2  & \phs 77      &   \phn2   &  16   &  & 0.84 & $-$47     &  56   \\
      19 &  $\rho$ Oph                                 & \phn41   & 1.6  &  0.2  & $-$21        &   \phn2   &  13   &  & 0.76 & \phs\phn7 &  29   \\
      20 &   IRAS 16293                                & \phn\phn7 & 0.6 &  0.2  & \phs 90      &   \phn3   &  25   &  & 0.67 & $-$65     &  26   \\
      21 &      NGC 6334\,V\footnote{Objects with angle dispersion larger than 30 degrees are not included in the angle analysis.}%
      \setcounter{fn}{8}                            & \phn\phn8 & 0.3 &  0.2  & \phs 90      &   \phn3   &  34   &  & 0.93 & $-$53     &  37   \\
      22 &      NGC 6334\,A                            & \phn49    & 1.3 &  0.2  & \phs 69      &   \phn2   & 18 &  & 0.89 & $-$\phn3  &  71   \\
      23 &      NGC 6334\,I                            & \phn54    & 0.9 &  0.2  & \phs 41      &   \phn2   & 19    &  & 0.51 & \phs11    &  30   \\
      24 &           W33 C$^\alph{fn}$                  & \phn29    & 0.4  &  0.2  & \phs 48      &   \phn2   & 37   &  & 0.88 & \phs82    &  35   \\
      25 &           W33 A                             & \phn19    & 0.9  &  0.2  & $-$53        &   \phn2   &  13  &  & 0.90 & $-$49     & \phn4 \\
      26 &            M17                              & 127       & 0.9  &  0.2  & $-$13        &   \phn2  &  27   &  & 0.59 & $-$10     & \phn3 \\
      27 &      W43-MM1                                & \phn\phn4 & 1.5  &  0.3  & \phs\phn 9   &   \phn4   &  22   &  & 0.88 & \phs34    &  25   \\
      28 &         G34$^\alph{fn}$                      & \phn44    & 0.5  &  0.2  & $-$66        &   \phn2   & 41   &  & 0.94 & \phs72    &  42   \\
      29 &           W49 A                             & \phn32    & 0.6  &  0.2  & \phs 59      &   \phn2   &  29   &  & 0.87 & $-$80     &  41   \\
      30 &           W51 A                  & 109       & 0.5  &  0.2  & \phs 44      &   \phn2   &  28   &  & 0.71 & $-$78     &  58   \\
      31 &            W75 N$^\alph{fn}$                & \phn\phn9 & 0.3  &  0.2  & \phs 44      &       \, 3
      & 47 &  & 0.98 & \phs76    &  33   \\
      32 &        DR21                                 & 142 & 1.2  &  0.2  & \phs\phn 5   &   \phn2   &  20   &  & 0.26 & \phs\phn0 & \phn5 \\
      \hline                                            
    \end{tabular}                                       
  \end{minipage}                                        
\end{table*}                                            
                                                        

\section{Analysis}\label{form}

\subsection{Projection Effects}

\begin{figure}
\includegraphics[width=80mm]{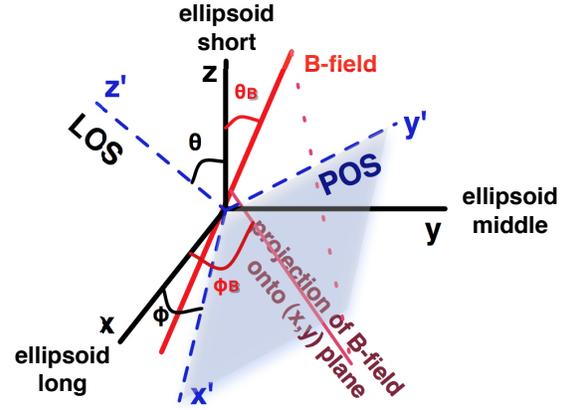}
\caption{Solid black lines: native cloud ellipsoid coordinate system
  $(x,y,z)$. Dashed blue lines: observation coordinate system
  $(x^\prime,y^\prime,z^\prime)$. The LOS is along the
  $z^\prime$ axis, and the POS is the
  $x^\prime$-$y^\prime$ plane. The $y^\prime$ axis represents the
  direction of the projection of the shortest ellipsoid axis onto the
  POS. The directions of the magnetic field and of its
  projection onto the $x$-$y$ plane are shown in red.\label{FIGURE0}}
\end{figure}

Let us consider a triaxial ellipsoid model molecular cloud with 
semi-axes $a \geq b \geq c$, and 
axial ratios 
\begin{equation}
\zeta = \frac{b}{b} 
\end{equation}
(middle-to-long axis ratio) and 
\begin{equation}
\xi = \frac{c}{a}
\end{equation}
(short-to-long axis ratio). When observed, the image of the molecular cloud appears
on the POS as an ellipse. The observable quantity related to the cloud
shape is the ellipse aspect ratio, $q$. Let us additionally assume that
the molecular cloud is threaded by a magnetic field, the mean
direction of which forms a polar angle $\theta_B$ with the short axis
of the cloud ellipsoid, and an azimuthal angle $\phi_B$ with the long
axis of the cloud ellipsoid (see Fig. \ref{FIGURE0}). Through
polarimetry observations, only the direction of the projection of the
field on the POS can be measured, and the associated observable
quantity is the angle $\lambda$ between the projected field and the
minor axis of the POS cloud ellipse.

The observables $q$ and $\lambda$ can be calculated as a function of
$\zeta$, $\xi$, $\theta_B$, $\phi_B$ and of the orientation angles of
the observer's LOS, $\theta$ and $\phi$. This calculation is
described in detail in Appendix \ref{approj}.

However, the orientation of the LOS is unknown --- in the
absence of biases, the LOS orientation can be treated as
uniformly distributed among different directions with respect to the
observer. As a result, it is not generally possible to de-project the
intrinsic cloud shape and magnetic field orientation for any single
object. Instead, a statistical treatment must be used: the preferred
intrinsic cloud shapes and magnetic field orientations in nature can
only be obtained by observing a large number of clouds, and comparing
the distribution of observables with expectations based on different
intrinsic shapes and field orientations. In this work, we employ such
a statistical analysis described below.

\subsection{Statistical Analysis}\label{san}

As a first proof-of-concept, we test for the presence of possible
degeneracies in the distribution of the observables ($q, \lambda$)
which may limit the value of this analysis. The simplest tests we can perform to verify that the data of Fig. \ref{FIGURE1} have nontrivial information content derive from the immediate observation that almost all datapoints lie below the diagonal extending from $(0.0,0)$ to $(1.0,90)$. We evaluate the probability to obtain this configuration by chance. 

First, we consider the case in which both aspect ratios and angles are randomly drawn from uniform distributions. Under this assumption, the desired probabilities can be obtained analytically: the diagonal splits the parameter space in two parts of equal area. Therefore, the probability for any single point to lie above or below the diagonal is $0.5$, and the outcome of multiple draws obeys the binomial distribution with probability for a positive outcome equal to $0.5$. If we consider all 32 datapoints (including the ones that did not survive our quality cuts), then we have 2/32 points above the diagonal. The probability to obtain the observed outcome or one which is {\em even more biased} towards points below the diagonal is the sum of the probabilities to obtain 0/32, 1/32 or 2/32 points above the diagonal: 
\begin{equation}
P({\rm \# \,\, above \,\, diagonal\leq 2}) = 
\sum_{i=0}^2 
\left(\!\!\begin{array}{c}32\\i\end{array}\!\!\right)0.5^i0.5^{32-i} = 1.2 \times 10^{-7}
\end{equation}
If we consider only the 24 datapoints that survived our quality cuts, then we have 1/24 points above the diagonal, and the associated probability to have obtained such a result by chance is 
\begin{equation}
P({\rm \# \,\, above \,\, diagonal\leq 1}) = 
\sum_{i=0}^1 
\left(\!\!\begin{array}{c}24\\i\end{array}\!\!\right)0.5^i0.5^{24-i} = 1.2 \times 10^{-7}
\end{equation}
with the result being numerically approximately equal as in the previous case. This configuration is extremely unlikely to have been obtained by chance from a uniformly distributed parameter space. 

Next, we consider the case in which the aspect ratio distribution in nature is identical to the observed aspect ratio distribution, but angles are randomly drawn from a uniform distribution. For each observed aspect ratio, we randomly select an angle $\lambda$, and we calculate the probability that the number of 
points below the diagonal is  $\leq 2$ for the 32 datapoints and $\leq 1$ for the 24 datapoints surviving the cuts. We find that this probability is equal to $0.98\%$ and $1.3\%$ respectively. Although more likely than before, this scenario also has a very small probability of occurring. 

\begin{figure*}
\includegraphics[width=160mm]{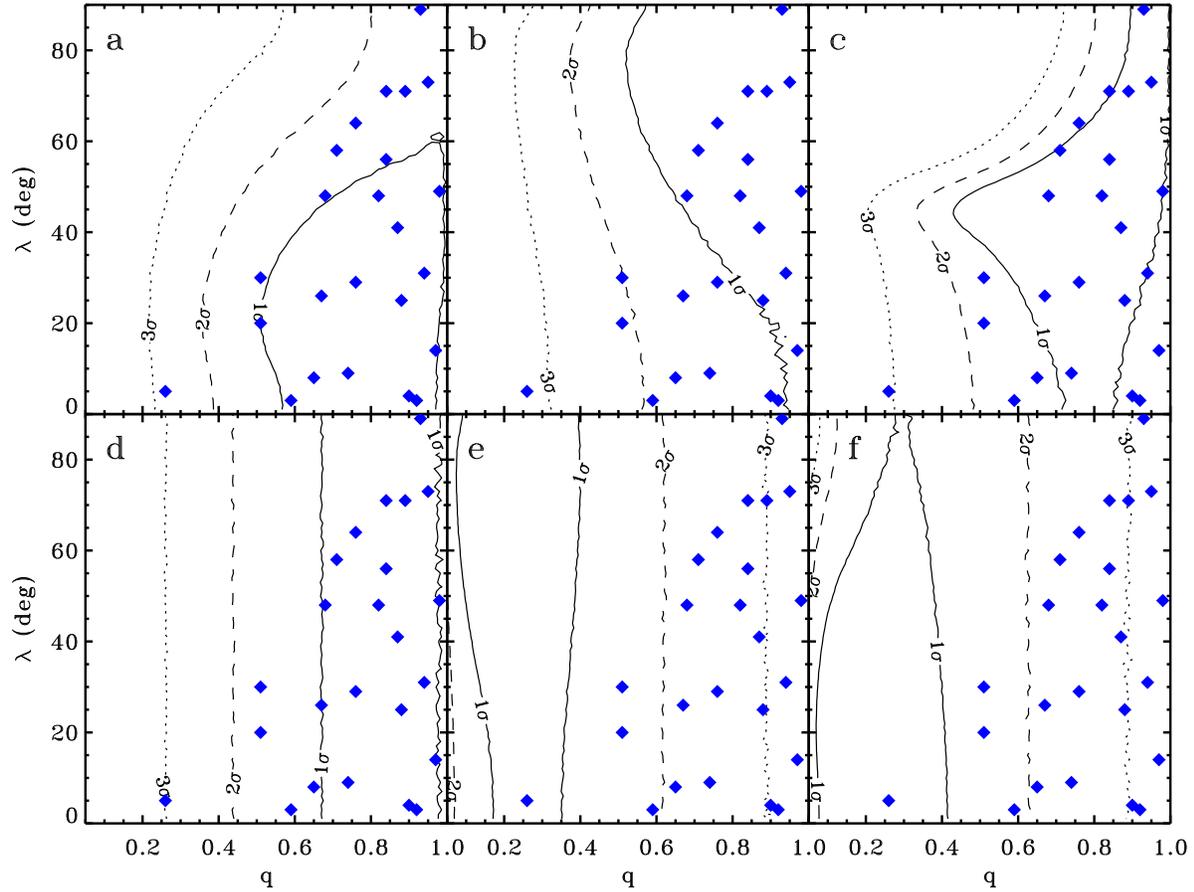}
\caption{Contour plots of the joint probability density function of
  the observables $q$ and $\lambda$ obtained by convolving different
  intrinsic distributions of cloud shapes and magnetic field
  orientations with random LOS. The solid, dashed, and
  dotted lines correspond to the $1, 2$, and $3\sigma$ contours,
  respectively. The blue data points correspond to our 24 data points
  surviving the quality cuts. Panel a: oblate clouds, magnetic field
  preferentially oriented parallel to short cloud axis; panel b:
  oblate clouds, magnetic field preferentially oriented parallel to
  the long cloud axis; panel c: oblate clouds, magnetic field
  preferentially oriented at a 45 degree angle with the short cloud
  axis; panel d: oblate clouds, uniformly distributed magnetic field
  orientations; panel e: prolate clouds, magnetic field preferentially
  oriented parallel to long cloud axis; paned f: prolate clouds,
  magnetic field preferentially oriented parallel to short cloud
  axis. The parameters of the intrinisc distributions resulting in
  each PDF are given in Table \ref{pdftable}.
\label{6pdfs}}
\end{figure*}

A more sophisticated version of these qualitative arguments can be made  starting from
characteristic limiting cases for the distribution of intrinsic cloud
features (shapes and magnetic field orientations). For each such case  we compute, by
performing ``mock observations'' along randomly selected and uniformly
distributed LOS, the joint probability density function
(PDF) of our two observables, the elongation $q$ and the projected
field - short ellipse axis angle $\lambda$. The results are shown in
Fig.~\ref{6pdfs}, overplotted with the 24 observed points that survive
the $\lambda$ dispersion cut described in \S \ref{obs}. The
parameters of the intrinsic shape and field orientation distributions
that lead to each PDF are given in Table \ref{pdftable}. These
parameters are: in the case of the intrinsic shape distribtion, the
cloud ellipsoid axial ratios $(\zeta_{\rm max}, \xi_{\rm max})$ of the
most probable shape (the values that maximize the PDF of axial
ratios); and in the case of the field orientation distribution, the
most probable cosine of the field orientation polar angle $\cos
\theta_{B, \rm max}$, and the standard deviation of the cosine of the
field polar angle, $\sigma(\cos \theta_B)$. The exact functional forms
of the shape and field orientation distributions used are discussed in
Appendices \ref{modlog} and \ref{delta}.

\begin{table*}
\begin{center}
\begin{tabular}{cccccc}
\hline
Fig. \ref{6pdfs}& $\zeta_{\rm max}$  & $\xi_{\rm max}$ & $\cos\theta_{B\rm max}$ & $\sigma (\cos \theta_B)$ & qualitative \\
panel& & & & & behavior\\
\hline
a&0.999 & 0.58 & 0.91 & 0.21& oblate, B $\parallel$ short axis\\
b&0.999 & 0.53 & 0.19 & 0.19 &oblate, B $\perp$ short axis\\
c&0.999 & 0.62 & 0.70 & 0.04 &oblate, B at $45^\circ$ with short axis\\
d&1.00 & 0.62 & \multicolumn{2}{c}{uniform in $\cos\theta_B$}& oblate, B random\\
e&0.29&0.08&0.19&0.19&prolate, B$\perp$ short axis\\
f&0.29&0.08&0.91&0.16&prolate, B$\parallel$ short axis\\
\hline
\end{tabular}
\end{center}
\caption{Parameters of the intrinsic distributions giving rise to the PDF in each panel of Fig. \ref{6pdfs}. \label{pdftable}}
\end{table*}

The qualitative distribution of the data points on the $q$-$\lambda$
plane forms a triangular shape, with the lower-right part of the plot
(almost circular cloud images, small angles of the magnetic field with
the short ellipse axis) more populated than the upper left part of the
plot (very elongated cloud images, large angles of the magnetic field
with the short ellipse axis). In general, there are many fewer very
elongated POS cloud ellipses than circular-looking ellipses, which
qualitatively indicates oblate intrinsic shapes (since there are many
more LOS that will yield a circularly-looking POS ellipse
for a disk-like cloud than for a cigar-like cloud). Additionally,
large angles with the short ellipse axis are only encountered for
almost circular cloud images (giving rise to the triangular
distribution of the data on the observables plane), which
qualitatively hints to an orientation of the magnetic field at small
angles with the short cloud axis (so that when the cloud is viewed
edge-on, the observations yield an elongated POS ellipse and a small
POS field angle with the short ellipse axis, cf.\ Fig.\ \ref{dr21},
while when viewed face-on the cloud looks circular and a larger
variety of angles of the POS field with the short ellipse axis are
possible, cf.\ Fig.\ \ref{g34}).

The different PDFs plotted in Fig. \ref{6pdfs} demonstrate how the
behavior of the PDF of the observables responds to changes in the
underlying distributions of shapes and field orientation. For oblate
shapes, the contours of the joint PDF for $\lambda$ and $q$ for
different field orientations are shown in panels a--d. As expected,
when the field is oriented parallel to the short cloud axis, the
$\lambda$-$q$ PDF acquires a roughly triangular shape preferentially
populating the high-$q$ low-$\lambda$ corner (panel a), while for a
field oriented perpendicular to the short cloud axis the PDF is also
roughly triangular, however now preferentially populating the high-$q$
high-$\lambda$ corner (panel b). For the intermediate situation of a
field oriented at $45^\circ$ from the short cloud axis, the PDF forms
a ``spike'' at the mid-$q$ mid-$\lambda$ part of the plot (panel
c). Finally, for a random distribution of field orientations (panel
d), the contours are parallel to the $\lambda$ axis, with low $q$
values preferred. PDFs resulting from prolate intrinsic cloud shapes
are shown in panels e and f, with panel e corresponding to a magnetic
field oriented along the long axis (filamentary clouds formed along
the magnetic field), and panel e corresponding to a magnetic field
oriented along the short axis (filamentary clouds perpendicular to the
magnetic field). In both cases, low values of $q$ are preferred (a
situation not seen in the data). The PDF of panel e preferenially
populates the high-$\lambda$ part of the plane, while the PDF of panel
f preferentially populates the low-$\lambda$ region.

We can therefore see that there is enough qualitative and quantitative
variation between these PDFs so that, with a sufficiently large
dataset, the effect of the random orientation angles can be overcome
and we can draw conclusions regarding the underlying
distributions of cloud shapes and magnetic field orientations in
nature.

In order to determine which such intrinsic distributions of cloud
parameters fit our data best, we bin the $\lambda$-$q$ parameter space
in $5\times5$ bins. For each bin, we count the number of observed data
points that fall within the bin limits.  We parameterize the intrinsic
shapes distribution as described in Appendix \ref{modlog}, using a
bi-parametric joint distribution in $\zeta$ and $\xi$ of non-zero
spread, uniquely defined by the values of $\zeta$ and $\xi$ where the
probability density becomes maximum, $\zeta_{\rm max}$ and $\xi_{\rm
  max}$. We parameterize the intrinsic distribution of magnetic fields
orientations as described in Appendix \ref{delta}, using a
bi-parametric distribution\footnote{Since a likelihood analysis on the
  shapes alone (see discussion in Appendix \ref{modlog}) indicates
  that the distribution of cloud shapes is stronly peaked on the
  $\zeta = 1$ axis, implying that the long and middle axes of the cloud
  ellipsoid are very close to being equal, the $\phi_B$ angles are
  physically degenerate, and the only quantity of interest determining
  the orientation of the magnetic field is the angle with the shortest
  axis of the cloud ellipsoid, $\theta_B$. } in $\cos \theta_B$,
uniquely defined by the value of $\cos\theta_B$ where the probability
density becomes maximum, $\cos\theta_{B\rm max}$, and the standard
deviation of the distribution, $\sigma(\cos \theta_B)$. Note that we
wish to know the distribution of $\cos \theta_B$ rather than that of
$\theta_B$, since the quantity of interest is the fraction of
LOS per solid angle in any given direction $\theta_B$,
which is expressed by $\cos \theta_B$ (see also discussion in Appendix
\ref{apstat}).  Assuming uniformly distributed LOS, we
determine the joint PDF for $q$, $\lambda$, and the expected counts
for 24 observations in each bin, as well as the spread in the expected
counts.  We perform a weighted least-squares analysis to find the
best-fit parameters $\zeta_{\rm max}, \xi_{\rm max}, \cos\theta_{B\rm
  max}, \sigma(\cos\theta_B)$, using the inverse square of the spread
in the expected counts as our weight. The details of this statistical
analysis are discussed in Appendix \ref{apstat}.  
Because of the nonlinearity of the problem, there are no applicable
analytic solutions to the minimization of the weighted sum of squared residuals. 
For this reason, we have used a well-tested, Monte-Carlo--based approach, 
the ``Simulated Annealing'' algorithm (Corana et al.\ 1987)
to  sample the parameter space and identifying the
location of the absolute minimum of the weighted sum of 
squared residuals.

\section{Results}\label{res}

\begin{figure}
\includegraphics[width=80mm]{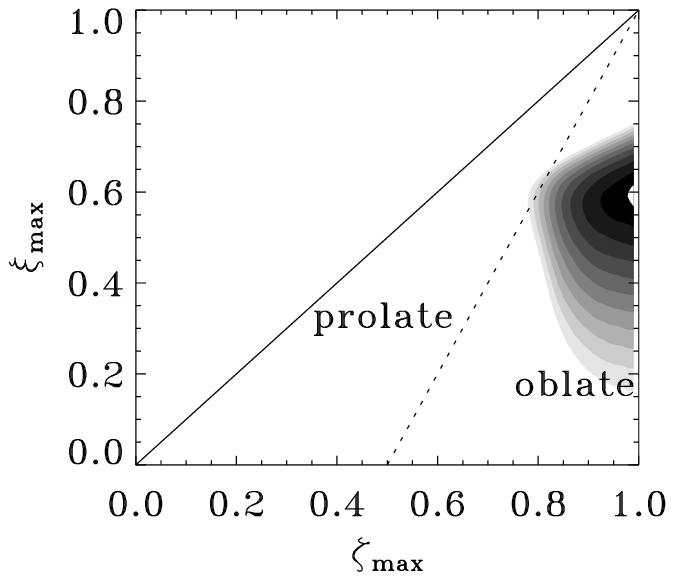}
\includegraphics[width=80mm]{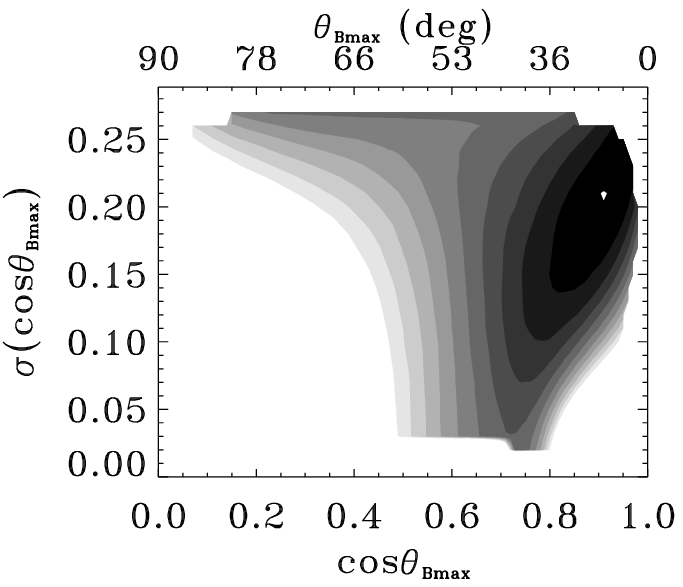}
\caption{Contours of the weighted sum of squared residuals, $S_{\rm
    weighted}$, for the $(\zeta_{\rm max},\xi_{\rm max})$ parameters
  (upper panel) and the $(\cos\theta_{B,\rm max},
  \sigma(\cos\theta_B)$ parameters (lower panel). The upper axis in
  the lower panel shows the most-probable orientation in degrees from
  the short cloud ellipsoid axis, corresponding to the value of $\cos
  \theta_{B,\rm max}$ shown in the lower axis. The two parameters {\em
    not shown} in each plot are kept fixed at their least-squares
  values. The color scale corresponds to values of $S_{\rm weighted}$
  from $9.7$ to $25$, with the contours spaced by factors of $1.1$ (so
  the suppressed $z$-axis in this plot which is visualized by the
  contours is in logarithmic scale).  The minimum is indicated in each
  case by the white point within the back region. The final contour
  corresponds to the $S_{\rm weighted}$ value that it typically
  yielded by 24 observations drawn from the best-fit intrinsic
  distributions due to random fluctuations, as calibrated by
  Monte-Carlo simulations.  The solid line in the upper panel
  separates the domain of allowed values ($\zeta_0\geq \xi_0$) from
  the rest of the plane, while the dotted line separates mostly oblate
  [$\zeta_{\rm max} \geq 0.5(1+\xi_{\rm max})$] from mostly prolate
  [$\zeta_{\rm max} < 0.5(1+\xi_{\rm max})$] most-probable
  ellipsoids. \label{CONTOURS1}}
\end{figure}

\begin{figure}
\includegraphics[width=80mm]{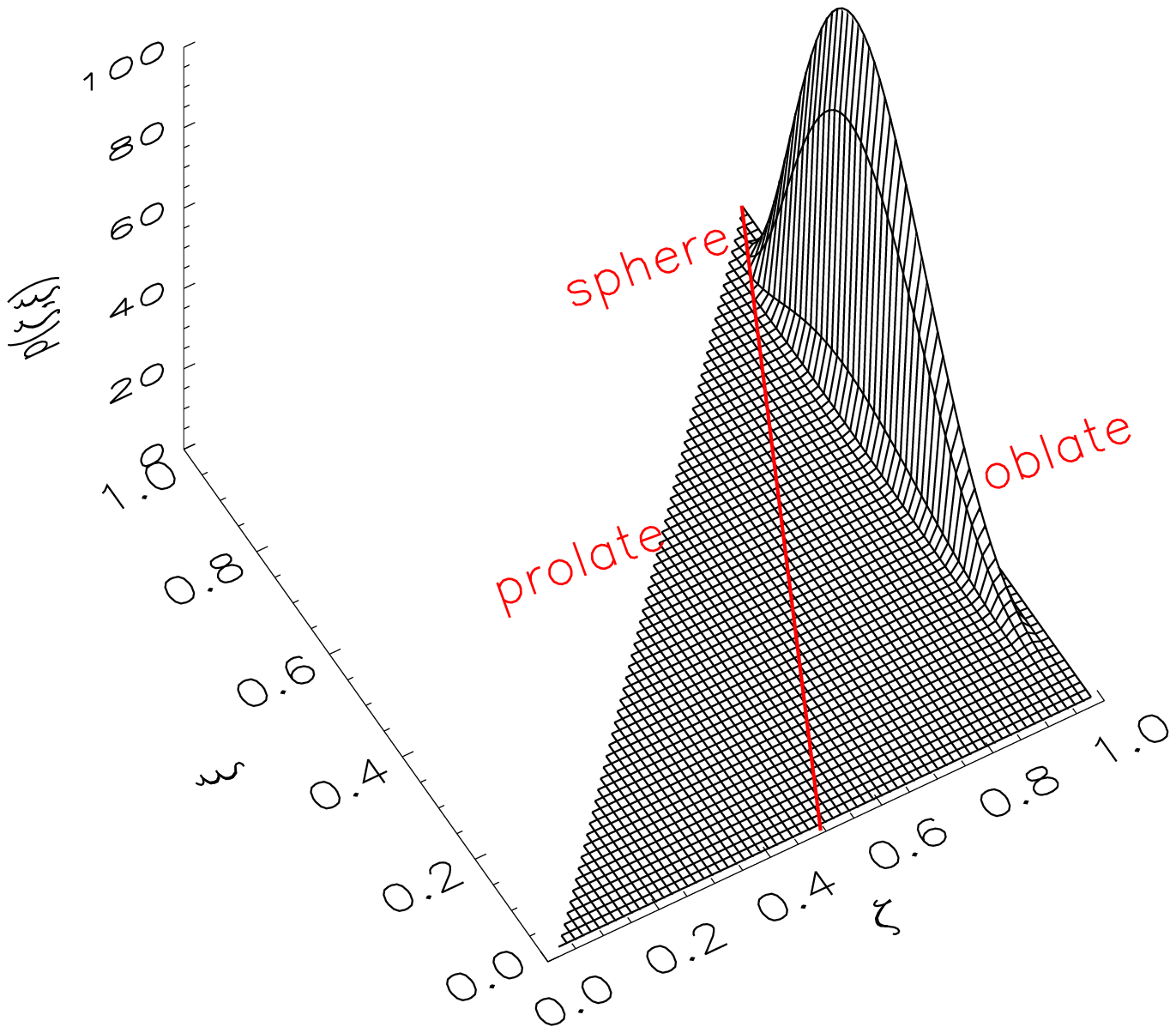}
\includegraphics[width=80mm]{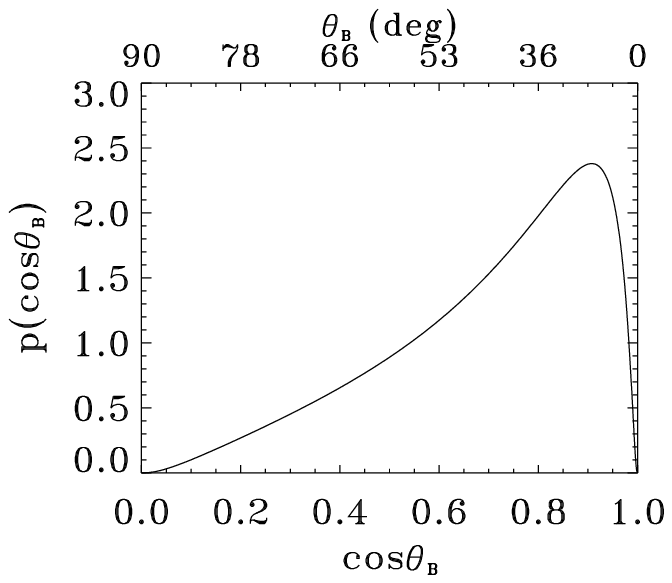}
\caption{Best-fit intrinsic distributions of cloud axial ratios (upper
  panel) and of magnetic field orientation angles (lower
  panel). In the upper panel, the red line extending from ($\zeta,\xi$)=(1,1) to 
($\zeta,\xi$)=(0.5,0) separates the $p(\zeta,\xi)=0$ plane 
in two sections: ellipsoids with axis ratios to the left of this line are mostly prolate, and 
ellipsoids to the right of the line are mostly oblate. Infinitesimally thin disks have 
($\zeta,\xi$)=(1,0); infinitesimally thin cigars have ($\zeta,\xi$)=(0,0); and perfect spheres
have ($\zeta,\xi$)=(1,1).  \label{DISTS1}}
\end{figure}

The results of our weighted least-squares analysis are shown in
Fig. \ref{CONTOURS1}. The upper panel shows contours of weighted
squared residuals, $S_{\rm weighted}$, as defined in Appendix
\ref{apstat}, on the $\zeta_{\rm max}$-$\xi_{\rm max}$ plane, with
$\cos \theta_{B \rm max}$ and $\sigma (\cos \theta_B)$ fixed at their
best-fit values.
The lower panel shows contours of $S_{\rm weighted}$ on the
$\cos\theta_{B,\rm max}$-$\sigma(\cos\theta_B)$ plane, with
$\zeta_{\rm max}$ and $\xi_{\rm max}$ fixed at their best-fit values.
The allowable values of $\sigma(\cos \theta_B)$ vary between 0
($\delta$-function of B-field orientations) and $0.289$ (uniform
distribution of B-field orientations with all orientations equally
probable, see Appendix \ref{apstat}).  The color scale corresponds to
values of $S_{\rm weighted}$ from $9.7$ to $25$, with the contours
spaced by factors of $1.1$ (so the suppressed $z$-axis in this plot
which is visualized by the contours is in logarithmic scale). The
location of the minimum is indicated in each case by the white point
within the back region. The outermost contour corresponds to the
$S_{\rm weighted}$ value that is typically yielded by 24 observations
drawn from the best-fit intrinsic distributions due to random
fluctuations, as calibrated by Monte-Carlo simulations.
 
As expected from the qualitative arguments outlined above, oblate
cloud shapes are preferred. In the case of the magnetic field
orientation, there is a well-defined broad minimum in the weighted
least-squares analysis, corresponding to a clear preference for a
small angle from the short cloud ellipsoid axis and a moderate spread
of angle about that. However, given the small number of data, the
power of the test to reject models is limited, and a large fraction of
the $\cos\theta_{B\rm max}$-$\sigma(\cos\theta_B)$ plane (excluding,
however, distributions strongly peaked at large angles from the
shortest cloud axis) could be consistent with the presently available
data.

The best-fit values of the parameters are shown in the first line of
Table \ref{pdftable}, and the resulting PDF of our observables, $q$
and $\lambda$, is shown in panel a of Fig. \ref{6pdfs}. The best-fit
distribution of intrinsic shapes (axial ratios of the model cloud
ellipsoids) and magnetic field orientation angles are shown in the
upper and lower panels of Fig. \ref{DISTS1}, respectively.  The most
probable shape is a moderately thick oblate disk. The best-fit shapes
distribution is very strongly peaked close to $\zeta \approx 1.0$ (so
clouds are likely to have very small degrees of triaxiality). On the
other hand, the distribution is very spread out in $\xi$, implying
that many different disk thicknesses are possible. Although the most
probable shape is a relatively thick disk, this does not mean that all
clouds are thick disks. Much thinner disks [closer to ($\zeta,\xi$)=(1,0)], as well as very thick
clouds [closer to ($\zeta,\xi$)=(1,1), labeled ``sphere'' on the $p(\zeta,\xi)=0$ plane], are also frequently encountered although not equally common in this distribution.

The most probable magnetic field orientation has a small offset
($\theta_{B0}\sim 24^\circ)$ from the short ellipsoid axis, and the
standard deviation in $\cos\theta_B$ of $\sim 0.21$. These results are robust against the removal of any outliers from our dataset of 24 observations. For example, when repeating the analysis without our most elongated cloud (DR 21), the best-fit parameters of the shape distribution change from $(\zeta_{\rm max},\xi_{\rm max})=(0.999,0.58)$ to $(\zeta_{\rm max},\xi_{\rm max})=(0.986,0.64)$, while the best-fit parameters of the $B-field$ orientations distribution change from $[\cos \theta_{V \rm max}, \sigma(\cos \theta_B)]=(0.91,0.21)$ to  $[\cos \theta_{V \rm max}, \sigma(\cos \theta_B)]=(0.86,0.19)$ - a change smaller than $10\%$ in any one of the parameters, while the qualitative behavior of the underlying distributions remains unchanged. This result is not surprising: in the upper-left panel of Fig. \ref{pdftable} DR 21 is outside the 2$\sigma$ contour of the best-fit joint PDF for $q$ and $\lambda$, and barely within the $3\sigma$ contour. 

As indicated by the broadness of the minimum of $S_{\rm
  weighted}$ as a function of our fitted parameters, the best-fit
intrinsic distributions of shapes and field orientations are not
unique in their ability to yield $q$-$\lambda$ PDFs that are
acceptable representations of the data used in this analysis. However,
both the qualitative arguments presented in \S \ref{san} as well as
the location of the minimum of $S_{\rm weighted}$ indicate that oblate
shapes and mean magnetic field orientations with small deviations from
the shortest axis of the cloud are preferred.

\section{Discussion}\label{disc}

In this paper, we have presented a new method for assessing
quantitatively the intrinsic shapes and the orientations of the mean,
ordered component of the magnetic field of molecular clouds, using
observations of the apparent aspect ratios $q$ and angles $\lambda$
between the apparent mean magnetic field and the apparent short axis
of the POS cloud images.  Under the assumption that the LOS
towards the observed clouds have random orientations with respect to
the principal cloud axes, we have explored different statistical methods to 
evaluate the consistency of various models for the intrinsic shapes and 
intrinsic magnetic field orientations of clouds with our data. We have used
data from 24 molecular clouds obtained through $350\,\mu$m
observations with the Hertz polarimeter
(Dotson et al.\ 2009). 

Based on our data sample we can exclude certain simple scenarios with high 
confidence. A scenario in which both cloud aspect ratios and magnetic 
field orientation angles are randomly drawn from uniform distributions 
can be excluded at the $10^{-7}$ level. A less restrictive scenario, in which 
the distribution of cloud aspect ratios is identical to the one observed 
in the data but magnetic field orientations are drawn randomly from a 
uniform distribution, is also excluded at the $10^{-2}$ level. 

In order to examine more general cases, 
we have employed a weighted least-squares
analysis to derive the best-fit intrinsic shapes and intrinsic
magnetic field orientations of the clouds in our sample. We have found that the most probable intrinsic
shape is a thick (short to long axis ratio $\sim 0.6$) oblate disk
with a negligible degree of triaxiality (middle to long axis ratio
$\sim 0.99$); the best-fit distribution of cloud thicknesses was found
to be broad, so both thin and thick clouds are frequently encountered
under this distribution.  The most likely orientation of the magnetic
field is close to the shortest axis of the cloud ($\sim
24^\circ$ offset toward the middle/long axis). The best-fit
distribution of magnetic field orientations is clearly peaked around
this value, but also features long tails.  We have found that the
best-fit distribution of shapes and orientations, when convolved with
a random LOS distribution yields a distribution of
observables which is in good agreement with the data. Our results are robust against the removal of any single point in our dataset (for example, if we repeat the analysis without our most elongated cloud, DR 21, the change in our best-fit parameters is smaller than $10\%$). 

These results are in agreement with the qualitative trend seen in
the data, where clouds with small apparent aspect ratio (apparently
elongated clouds) have small angles between the mean projected
magnetic field direction and the short axis of the cloud ellipse;
clouds which are apparently circular on the other hand can have a
large range of angles between mean magnetic field and short
axis. Finally, apparently elongated clouds with large angles between
mean field direction and short cloud axis are not seen in the data. As far as
the distribution of apparent cloud shapes is concerned, more clouds
are seen to have large aspect ratios (and are thus close to circular)
while only a few clouds are apparently very elongated.

Indeed, clouds which are intrinsically disks with the magnetic field
direction close to that of the shortest cloud axis would exhibit
similar properties when viewed through random LOS. More
LOS yield a disk seen almost face-on rather than edge-on,
so most clouds would look almost circular and only a few would appear
significantly elongated (in contrast to prolate clouds, most of which
would appear elongated, and randomly shaped clouds, which would show
no preference in apparent aspect ratio). In addition, when clouds are
seen almost edge-on and appear elongated, the magnetic field will be
aligned with the short axis of the projection of the cloud on the
POS. In contrast, when clouds are seen face-on, the magnetic field
projection could from any angle with respect to the short and long
axes of the cloud ellipse, which in reality are, in this case,
projections of the cloud ellipsoid middle and long axes.

Despite the preference of the data for such a scenario, and the clear
potential of this test for discriminating between scenarios for
intrinsic magnetic field orientations in molecular clouds and between
theories for molecular cloud dynamics, we have shown quantitatively
that a larger number of observations is needed for the test to
conclusively reject alternative configurations. With the present,
limited number of observed clouds, the distributions of LOS, magnetic field orientations, and shapes, are still sparsely
sampled, and as a result random fluctuations in this sampling allows
for a large number of parameters to yield acceptable representations
of the observed $q$ and $\lambda$. However, our results explicitly
demonstrate that there is no {\em instrinsic} degeneracy in the
distributions of $q$ and $\lambda$ yielded by different classes of
cloud shapes and magnetic field orientations in nature, so a large
number of $q$, $\lambda$ observations should allow conclusive tests
of different underlying distributions.

Our analysis has the advantage that it is not {\it a priori} tied to
any theory or prediction regarding the dynamical processes in
molecular clouds, but rather allows the data to pick freely the part
of the shapes/orientations parameter space that best fits the
observations. We have additionally tested that the location of the
best-fit $\zeta_{\rm max}$, $\xi_{\rm max}$, $\cos\theta_{B\rm max }$,
and the spread of the orientations distribution are robust with
respect to changes in the assumed functional form or the details of
the analysis (for example, performing a joint analysis for the shapes
and orientations, or performing an analysis for the shapes alone).
Finally, observational uncertainties in the measured quantities are
expected to have a limited effect in our analysis, as the binning of
the $(q,\lambda)$ parameter space involves bins that are typically
of the same order as or wider than such uncertainties.

The {\em apparent} orientation of the magnetic field in molecular
clouds and cloud cores has been studied in the past by various
authors. Kane et al.\ (1993), using 1.3\,mm polarization measurements,
claimed a strong preference for alignment of the polarized emission with  with the structure in deconvolved IRAS maps. Glenn et
al.\ (1999) used a sample of 7 elongated cloud cores selected to have
a polarization detection greater than $3\sigma$ at either 800\,$\mu$m
or 1.3\,mm. They found that the orientation of the apparent magnetic
field was random with respect to the apparent cloud axes. Vall\'{e}e
\& Bastien (1999), using 760\,$\mu$m observations, found the apparent
magnetic field direction in molecular cloud cores (intensity peaks within
clouds) to be parallel to the apparent minor axis in 3 out 10 cases.  These
studies all focused on polarization vectors measured at intensity
peaks, contrary to our own survey, where polarization is extensively
mapped throughout the clouds in our sample.  The intrinsic {\em
  shapes} of molecular clouds were discussed by Kerton et
al.\ 2003. Based on the Five College Radio Astronomy Observatory Outer
Galaxy Survey in CO emission, they studied a sample of 15,000
clouds. They concluded that the intrinsic shapes of these objects are
best described as intermediate between near-oblate and near-prolate
ellipsoids.

Our results have important implications concerning the dynamical
importance of magnetic fields in molecular clouds.  If magnetic fields
are dynamically unimportant compared to turbulent motions, the fields
are expected to be carried around with turbulent eddies in the cloud,
and show little correlation with the principal cloud axes. However,
our data show clear indications that such a correlation exists.  Our
results are also unfavorable for the scenario of helical magnetic
fields threading prolate clouds. Not only are the predicted $90^\circ$
flips in the polarization vectors not observed (Dotson et al.\ 2009),
but clouds in our sample are found to be oblate, rather than prolate.
This result is also consistent with the argument against prolate
cloud {\em cores}, which are also predicted in this scenario (Fiege \&
Pudritz 2000c), but are not observed in nature (Jones et al.\ 2001,
Jones \& Basu 2002, Tassis 2007).

If, on the other hand, magnetic fields are dynamically important,
cloud shapes should be close to oblate disks, and the magnetic field
should be closely aligned with the short axis of the cloud
ellipsoid. This scenario seems to be the one which yields the best
agreement with our data. Intrinsic cloud shapes are indeed very
strongly peaked around oblate disks. The best-fit magnetic field
orientation is close to the shortest cloud ellipsoid axis, with small
offsets toward the middle/long axes. However, more data are
required before a more robust quantitative statement can be made with
regard to both the details of the best-fit underlying magnetic fields
orientation distribution as well as the confidence with which
alternative scenarios can be rejected.

In addition to support from our statistical treatment, this picture is
also consistent with observations of individual well-studied systems
where projection effects are minimal. In these cases, the hourglass
morphology of the magnetic field, characteristic of dynamically
important magnetic fields, is seen in very different scales, from
clouds to dense cores (e.g., Schleuning 1998, in the case of the cloud
OMC-1; Girart, Rao \& Marrone 2006, in the case of protostellar core
NGC1333; Lai et al.\ 2002, in the case of massive core NGC2024).

We conclude by stressing that although our analysis has yielded
best-fit distributions that seem to {\em prefer} this portion of the
parameter space, the rejection power of our test, especially in the
case of magnetic field orientations, is limited given the present
sample size. Stronger constraints on the intrinsic statistics of
magnetic field orientations in nature will require polarimetry
observations in a larger number of clouds.

\section*{Acknowledgements}
KT thanks Shantanu Basu and Vasiliki Pavlidou for useful discussions,
and Nick Scoville, Paul Goldsmith and Telemachos Mouschovias for
feedback on the manuscript. We thank the referee, Carl Heiles, for 
insightful comments which helped us improve the manuscript. 
KT acknowledges support by NSF grants AST 02-06216 and AST02-39759, by
the NASA Theoretical Astrophysics Program grant NNG04G178G and by the
Kavli Institute for Cosmological Physics at the University of Chicago
through grants NSF PHY-0114422 and NSF PHY-0551142 and an endowment
from the Kavli Foundation and its founder Fred Kavli.  JEV
acknowledges support from NSF AST 05-40882 through the Caltech
Submillimeter Observatory. 
RH and LK acknowledge support from NSF grant AST 0505124.
Part of this work was carried
out at the Jet Propulsion Laboratory, California Institute
of Technology, under a contract with the National
Aeronautics and Space Administration.  \copyright 2008. All rights reserved. 

\begin{appendix}
\section{Projection Effects}\label{approj}

Let us consider a triaxial ellipsoid model molecular cloud of
semi-axes $a \geq b \geq c$, and a {\em native} system of coordinates
$(x,y,z)$ centered on the cloud and aligned with its axes. The
triaxial ellipsoid {\em surface} in this system obeys
\beq\label{ONE}
\frac{x^2}{a^2} + \frac{y^2}{b^2} + \frac{z^2}{c^2} = 1\,.
\eeq
The $x$, $y$, and $z$ axes are thus oriented along the longest,
middle, and shortest ellipsoid axes respectively. Let us additionally
define a second, {\em observation} system of coordinates,
$(x^\prime,y^\prime,z^\prime)$. In this system, the LOS is along the
$z^\prime$ axis, and consequently the $x^\prime$-$y^\prime$ plane is
the POS\@. The $x^\prime$ axis is in the $x$-$y$ plane, which implies
that $x^\prime\perp z$. The $y^\prime$ axis thus represents the
projection onto the POS of the $z$ axis, and consequently of the
shortest axis of the cloud ellipsoid. The orientation between the
native and observation systems of coordinates is given by the angles
$\theta$ and $\phi$ between the $z$ and $z^\prime$ axes and the $x$
and $x^\prime$ axes respectively (see Fig.\ \ref{FIGURE0}).

When observed, the image of the molecular cloud appears on the POS as
an ellipse. The properties of this projected, observed ellipse can be
calculated from the properties of the cloud ellipsoid and the
orientation of the observer's LOS\footnote{Note that the
  projection we are referring to here is not a ``cut'' of the cloud
  ellipsoid along some plane, but is rather the surface brightness
  along the observer's LOS  (and hence it represents an
  integral of the cloud density along the LOS)}(Binney
1985).  We define the axial ratios as $\zeta = b/a$ and $\xi = c/a$,
so $1\geq \zeta \geq \xi$.  Then, the POS isophotes of an
ellipsoid molecular cloud, the shape of which is described by
Eq.\ (\ref{ONE}), will be coaxial ellipses of aspect ratio
\begin{equation}\label{MOO}
1\geq q (\theta, \phi, \zeta, \xi) = 
\sqrt{\frac{A+C-\sqrt{(A-C)^2+B^2}}
{A+C+\sqrt{(A-C)^2+B^2}}}
\end{equation}
where 
\beq\label{THEA}
A \equiv \frac{\cos^2\theta}{\xi^2}
\left(\sin^2\phi +\frac{\cos^2\phi}{\zeta^2}\right)
+\frac{\sin^2\theta}{\zeta^2}\,,
\eeq

\beq\label{THEB}
B \equiv \cos \theta \sin 2\phi
\left(1-\frac{1}{\zeta^2}\right)\frac{1}{\xi^2}\,, 
\eeq

\beq\label{THEC}
C \equiv \left(\frac{\sin^2\phi}{\zeta^2} + \cos^2 \phi
\right) \frac{1}{\xi^2}\,.
\eeq

The orientation of the axes of the POS cloud ellipse with respect to
the POS axes ($x^\prime$ and $y^\prime$) is given by the angle $\psi$,
defined as
\beq\label{THEPSI} \psi =
\frac{1}{2}\arctan\left(\frac{B}{A-C}\right)\,.
\eeq
In Eq.\ (\ref{THEPSI}), $\psi$ can be the angle between either the
minor or the major axis of the POS ellipse with the $y^\prime$ axis
(the sum of the two angles is always equal to $90^\circ$, and $\psi$,
which can take values between $0$ and $45^\circ$, is always the
smaller of the two). Which one of the two angles is represented by
$\psi$ in each case is determined by the sign of the quantity
\begin{equation}\label{theu}
u=(A-C)\cos2\psi+B\sin2\psi\,.
\end{equation}
If $u\le 0$, $\psi$ represents the angle between the minor ellipse
axis and $y^\prime$.  If $u>0$, $\psi$ represents the angle between the
major axis and $y^\prime$. The sign of $\psi$ indicates the quadrant in
which the relevant axis (major or minor, depending on the sign of $u$)
resides. If $\psi\leq 0$ the relevant axis is located in the
first/third quadrant, while if $\psi >0$ the relevant axis is located
in the second/fourth quadrant.

Additionally, let us consider that the cloud is threaded by a
large-scale ordered magnetic field $\vec{B}$, which has some
orientation with respect to the cloud principal axes characterized by
a set of angles $(\theta_B, \phi_B)$; here, $\theta_B$ is the angle of
the magnetic field with the shortest cloud axis $z$ and $\phi_B$ is
the angle of its projection onto the plane $(x,y)$ defined by the
directions of the middle and longest axes (see
Fig.\ \ref{FIGURE0}). We are interested in the orientation of the POS
projection of $\vec{B}$, since this is what can be measured through
polarimetry observations. The components of $\vec{B}$ in the $(x,y,z)$
coordinate system are
\begin{equation}
\vec{B} = B\left(\begin{array}{l}\sin\theta_B\cos\phi_B
\\\sin \theta_B \sin \phi_B \\ \cos\theta_B \end{array}\right)\,.
\end{equation}
Performing the rotation to the $(x^\prime,y^\prime,z^\prime)$
coordinate system we find that the $x^\prime$ and $y^\prime$
components of $\vec{B}$ (the POS components) are
\begin{equation}
B_{x^\prime}=B \left[-\sin\phi\sin\theta_B\cos\phi_B + \cos\phi
  \sin\theta_B\sin\phi_B\right] 
\end{equation}
and
\begin{eqnarray}
B_{y^\prime}&=&B \left[-\cos\phi\cos\theta \sin\theta_B\cos\phi_B  \right. \nonumber
  \\ && \left. -\sin\phi\cos\theta
  \sin\theta_B\sin\phi_B +\sin\theta\cos\theta_B\right]\,.
\end{eqnarray}
The angle $\omega$ between the POS projection of $\vec{B}$ and the POS
projection of the shortest cloud axis (the $y^\prime$ axis) is given by
\begin{equation} \label{omegaeq}
\tan \omega =\frac
{\sin \phi \sin \theta_B \cos \phi_B - \cos \phi \sin \theta_B \sin \phi_B}
{\cos\!\theta\! \sin\!\theta_B\!\left(\! \cos \!\phi \!\cos\!\phi_B 
\!+\! \sin\! \phi\!\sin\!\phi_B\!\right)\!
- \!\sin\! \theta\! \cos\! \theta_B\!}\,.
\end{equation}
Note that, contrary to $\psi$, the angle $\omega$ is, by definition,
positive in the first-third quadrant, and negative in the
second-fourth quadrant, where $x^\prime$ and $y^\prime$ have opposite signs.

\begin{figure}
\includegraphics[width=80mm]{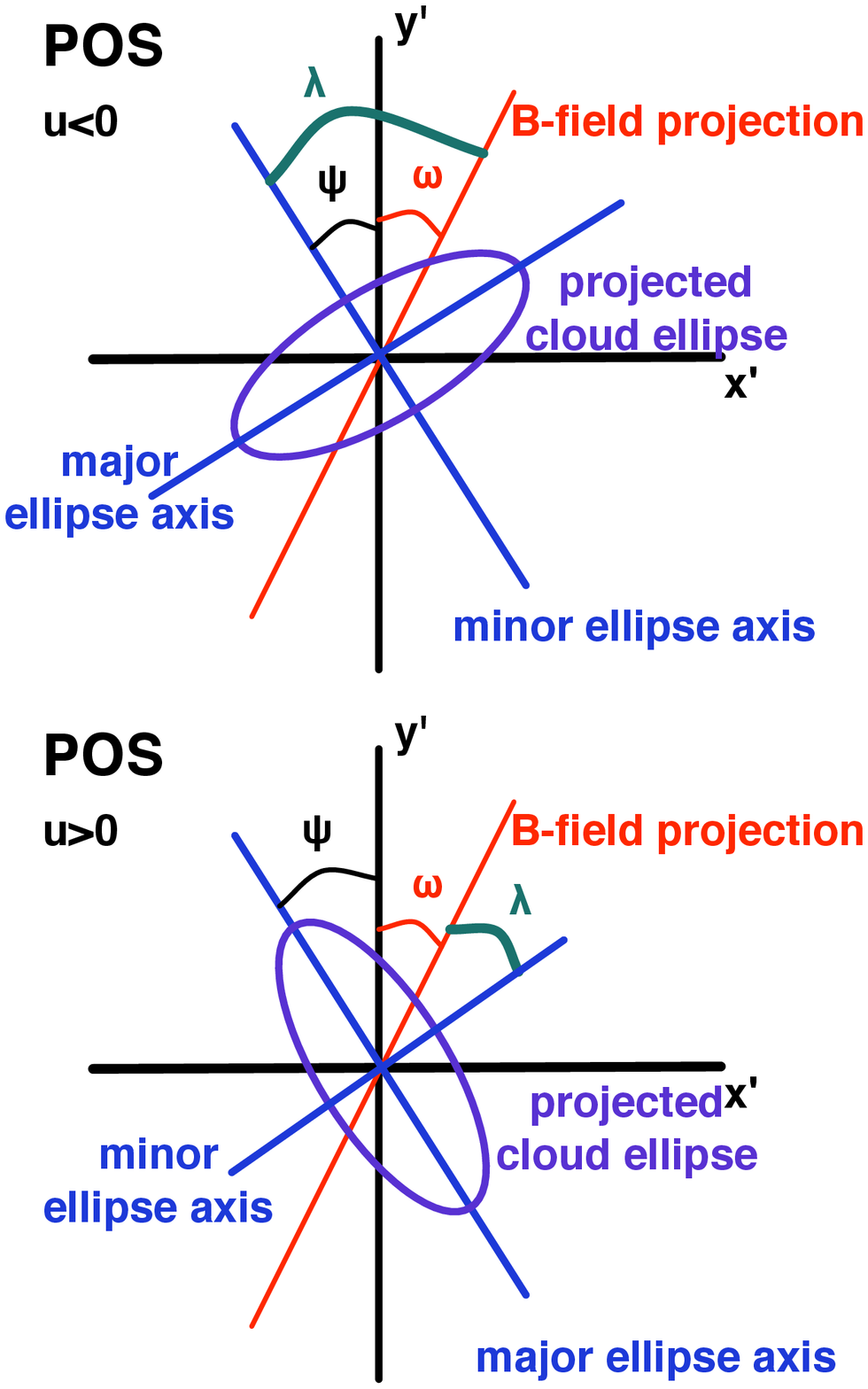}
\caption{POS angles between the POS projection of the
  magnetic field (red line), the POS projection of the shortest cloud
  ellipsoid axis (the $y^\prime$ axis), and the projected cloud ellipse axes
  (blue lines). Upper panel: $u<0$ ($\psi$ represents the angle
  between $y^\prime$ and the {\em minor} ellipse axis). Lower panel: $u>0$
  ($\psi$ represents the angle between $y^\prime$ and the {\em major}
  ellipse axis). The magnitude of the angle $\lambda$ is the
  observable angle between the POS cloud ellipse minor axis and the
  POS projection of the magnetic field. In this case, $\psi$ and
  $\omega$ are both positive, and $|\psi+\omega|<90^\circ$.
\label{FIGURE0b}}
\end{figure}

The angle that can be determined observationally, however, is not
$\omega$, but rather the magnitude of the angle $\lambda$ between the
POS projection of $\vec{B}$ and one axis (for definiteness we will use
the minor axis) of the POS cloud ellipse. The relation between
$\lambda$, $\omega$, and $\psi$ is different, depending on the value
of $u$ (see Fig.\ \ref{FIGURE0b}):
\begin{equation}\label{OINK}
\lambda = \left\{ \begin{array}{lr}
90^\circ- \left|
90^\circ -\frac{}{}|\psi+\omega| \right|,& u\leq 0 \\
\left|
90^\circ - \frac{}{}|\psi +\omega|\right|, & u>0 
\end{array}
\right.
\,.
\end{equation}

We have therefore shown that the two observable quantities $q$ and
$\lambda$ can be calculated using the formalism described above,
provided that we know:
\begin{itemize}
\item[(a)] the intrinsic shape of the cloud ellipsoid (i.e., its axial
  ratios $\zeta$ and $\xi$);
\item[(b)] the orientation of the ordered magnetic field with respect
  to the native coordinate system (i.e., the angles $\theta_B$ and
  $\phi_B$); and
\item[(c)]  the orientation of the LOS with respect to
  the native coordinate system (i.e., the angles $\theta$ and $\phi$).
\end{itemize}
Note that the angles $\theta_B$ and $\phi_B$ only enter the
calculation of $\lambda$, while $q$ is not dependent on them.
Because the LOS is random, $\lambda$ is not fixed, even if
the true magnetic field direction is fixed with respect to the true
cloud ellipsoid axes, {\em and} all molecular clouds have the exact
same intrinsic shape (see e.g., Basu 2000). Rather, there is a
distribution of such angles, which can be calculated analytically,
given our hypothesis for the orientation of the true mean magnetic
field in the clouds, as well as some knowledge of the distribution of
intrinsic cloud shapes in nature.

\section{Statistical Analysis}\label{apstat}

In this work, we use our set of observations of the aspect ratio $q$
of the POS cloud ellipse and the magnitude of the angle $\lambda$
between the POS projection of the magnetic field and the minor axis of
the POS cloud ellipse to constrain the intrinsic shapes and magnetic
field orientations in molecular clouds.  Unfortunately, it is not
possible to derive directly the intrinsic shape and the orientation of
the magnetic field in each cloud, as this would require knowledge of
the LOS orientation in each case. Instead, we employ a
statistical analysis, assuming that the distribution of LOS
orientations is uniform with respect to all directions (same fraction
of LOS per unit solid angle in all directions). In
practice, this means that the distribution of $\cos\theta$ is uniform
with values between $-1$ and $1$, and the distribution of $\phi$ is
also uniform with values between $0$ and $2\pi$.

Under this assumption, we will perform a weighted least-squares
analysis to deduce the best-fit probability density function of our
observables, $q$ and $\lambda$, and the associated distributions of
intrinsic shapes and orientations of the magnetic field with respect
to the principal cloud ellipsoid axes that generate it.

Because the magnetic field orientation does not enter the calculation
of $q$, it is also possible to use {\em only} the observations of $q$
to derive the best-fit distribution of the intrinsic
molecular cloud shapes (an analysis similar to that of Tassis 2007 in
the case of molecular cloud cores). 
The results obtained by this $q-$only analysis indicate that the
distribution of intrinsic shapes is strongly peaked on the $\zeta=1$
axis (largest and middle axes equal) so that all azimuthal magnetic field
angles $\phi_B$ are physically equivalent. For this reason, we will
treat the orientation angle $\phi_B$ as random and uniformly
distributed, and we will only attempt to determine the best-fit
distribution of $\cos\theta_B$, determining the most likely deviation
of the magnetic field orientation from the shortest cloud axis. Note
that the relevant quantity is $\cos\theta_B$, rather than $\theta_B$,
as we are not only interested in the most probable angle, but also on
the spread of such angles: one of the theoretical cases we would like
to test is that where all magnetic field orientations have an equal
probability of occurring.  This translates to a uniform distribution of
magnetic field orientations per solid angle around the cloud center,
which in turn is mathematically expressed as a uniform distribution in
$\cos\theta_B$. Thus, the interpretation of a best-fit distribution in
$\cos\theta_B$ is intuitively straight-forward: the standard deviation
corresponding to a $\delta$-function in preferred field orientation is
$\sigma(\cos\theta_B) = 0$, while the standard deviation corresponding
to a uniform orientation distribution is $\sigma(\cos\theta_B) =
\sqrt{\langle(\cos\theta_B)^2\rangle - \langle\cos\theta_B\rangle^2} =
\sqrt{1/3 - 1/4} = 0.289$. Moderately spread distributions will have a
standard deviation in between.

Our primary statistical analysis is performed under the following
assumptions.
\begin{itemize}
\item [i.]Molecular clouds can be described as triaxial ellipsoids of
  the form of Eq.\ (\ref{ONE}).
\item[ii.] The axial ratios $\zeta$ and $\xi$ of molecular cloud
  ellipsoids have an intrinsic distribution in nature, of the form
  described in Appendix \ref{modlog}, with two free parameters
  $\zeta_{\rm max}$ and $\xi_{\rm max}$ (the $\zeta$ and $\xi$ which
  maximize the distribution and correspond to the most probable
  shape).
\item[iii.] $\phi_B$ has a uniform distribution between $0$ and
  $2\pi$; $\cos\theta_B$ has an intrinsic distribution of the form
  described in Appendix \ref{delta}, with two free parameters
  $\cos\theta_{B,\rm max}$, and $\sigma(\cos\theta_B)$ (corresponding
  to the most probable orientation and the standard deviation of the
  distribution of $\cos\theta_B$).
\item[iv.] The orientation angles of the LOS with respect to the
  native coordinate system, $\theta$ and $\phi$, are random and all
  equally probable: the distribution of $\cos \theta$ is uniform with
  values between $-1$ and $1$, and the distribution of $\phi$ is
  uniform with values between $0$ and $360^\circ$.
\end{itemize}

Based on these assumptions, we calculate, for every set of
distribution parameters $\zeta_{\rm max}$, $\xi_{\rm max}$,
$\cos\theta_{B,\rm max}$, and $\sigma(\cos \theta_B)$, the expectation
value of the number of observed points in each bin on the
$q$-$\lambda$ plane, if 24 observations are performed with random
LOS; we also calculate the spread of the observed number of
points. We do so using the following Monte-Carlo procedure.

\begin{enumerate}
\item We randomly draw a pair of $\zeta, \xi$ and a $\cos\theta_B$ from their
  respective distributions, and a $\phi_B$ from a uniform distribution
  between 0 and $2\pi$.
\item We randomly draw a LOS orientation, by drawing a pair
  of $\theta$, $\phi$ from a uniform probability distribution (equal
  probability for any value of $\cos\theta$ or $\phi$ in the intervals
  $(0,1)$ and $(0,2\pi)$, respectively).
\item We use the values of $\zeta$, $\xi$, $\theta$, and $\phi$ to
  calculate $q$, $u$, and $\psi$ through Equations
  \ (\ref{MOO})-(\ref{theu})
\item We use the values of $\theta_B$, $\phi_B$, $\theta$, and $\phi$
  to calculate $\omega$ through Eq.\ (\ref{omegaeq}).
\item We use the values of $\omega$, $\psi$, and $u$ to calculate
  $\lambda$ through Eq.\ (\ref{OINK}).
\item We repeat steps (i)--(v) 24 times (as many as our observed
  clouds). For every bin in the $q$-$\lambda$ plane, we calculate
  the number of observations that fall within its limits.
\item We repeat step (vi) 10,000 times. Using the 10,000 mock sets of
  24 observations, we calculate for each bin $i$ the mean number of
  observations, $\langle N_i\rangle$ (which has to be $0\leq \langle
  N_i \rangle \leq 24$), and the sample variance of $N_i$,
  $\sigma_{Ni}^2 = \langle N_i^2 \rangle - \langle
  N_i\rangle^2$. Because we only use 10,000 mock observations and we
  use the inverse of $\sigma_{Ni}^2$ as a weight in our least-squares
  analysis, for the cases wherein our 10,000 trials we obtain
  $\sigma_{Ni}^2 = 0$ we use instead a floor value of $\sigma_{Ni}^2 =
  10^{-4}$
\end{enumerate}
 
We finally obtain the best-fit set of parameters $\zeta_{\rm max}$,
$\xi_{\rm max}$, $\cos\theta_{B,\rm max}$, and $\sigma(\cos \theta_B)$
by minimizing the weighted squared residuals function,
\begin{equation}
S_{\rm weighted}( \zeta_{\rm max},\xi_{\rm max}, \cos\theta_{B,\rm max}, \sigma(\cos \theta_B))  = \sum_{i=1}^{5\times5} \frac{N_{\rm data,i} - \langle N_i\rangle}{\sigma^2_{N_i}}
\,.
\end{equation}

\subsection{Modified Lognormal Distribution of Intrinsic Cloud Shapes}
\label{modlog}

We wish to use a distribution of finite width in each of the $\zeta,
\xi$ axes which is, however, smooth and has a relatively sharp
maximum. Such a distribution is a modified lognormal, which we
construct in the following way: we first seek an appropriate minimal
transformation which will transform the domain of $\zeta$ and $\xi$
from $(0,1)$ to $(-\infty,\infty)$; we then take the transformed
variable to follow a gaussian distribution.

An appropriate transformation is $\zeta\rightarrow x$ and $\xi
\rightarrow y$ where
\begin{equation}
x = \ln \frac{\zeta}{1-\zeta}\,,
\end{equation}
and 
\begin{equation}
y = \ln \frac{\xi}{\zeta - \xi}\,.
\end{equation}
It is easy to see that 
\[\lim_{\zeta \rightarrow 0} x = -\infty\,, \,\,\,
\lim_{\zeta \rightarrow 1} x = \infty\,,
\lim_{\xi \rightarrow 0} y = -\infty\,, \,\,\,
\lim_{\xi \rightarrow 1} y = \infty\,.
\]

If then the joint probability distribution function (PDF of $x$ and
$y$ is
\beq
p(x,y) = \frac{1}{2\pi\sigma_x\sigma_y}
\exp\left[-\frac{(x-x_0)^2}{2\sigma_x^2}\right]
\exp\left[-\frac{(y-y_0)^2}{2\sigma_y^2}\right]\,,
\eeq
then the joint PDF of $\zeta$ and $\xi$ will be
\begin{eqnarray}
p(\zeta,\xi)&=& \frac{1}{2\pi\sigma_x\sigma_y}
\exp\left[-\frac{(\ln\frac{\zeta}{1-\zeta}-x_0)^2}{2\sigma_x^2}\right]\nonumber
\\
&& \times \exp\left[-\frac{(\ln\frac{\xi}{\zeta-\xi}-y_0)^2}{2\sigma_y^2}\right]
\frac{1}{\xi(\zeta-\xi)(1-\zeta)}\,.
\end{eqnarray}

This is in principle a tetra-parametric distribution with parameters
$x_0$, $y_0$, $\sigma_x$, and $\sigma_y$. However, to reduce the
number of parameters entering the problem and the associated
computation time required for the analysis, while at the same time
allowing the expected amount of variation in molecular cloud shapes,
we fix the value of $\sigma_x$ and $\sigma_y$ to 0.2 and 0.7
respectively, determined from a likelihood analysis\footnote{For details on the methodology of this shapes-only likelihood analysis see Tassis 2007} of the aspect
ratios $q$ alone.

The values $\zeta = \zeta_{\rm max}$ and $\xi=\xi_{\rm max}$ that
maximize $p(\zeta,\xi)$ are given by the system of equations
\begin{eqnarray}
\partial p / \partial \xi &= &0\\
\partial p /\partial \zeta & = & 0
\end{eqnarray}
or, equivalently, 
\begin{eqnarray}
0&=& \sigma_y^2 \left(\ln \frac{\zeta}{1-\zeta} - x_0\right)(\zeta -
\xi)\nonumber \\
&& -\sigma_x^2\left(\ln \frac{\xi}{\zeta - \xi}
- y_0\right)(1-\zeta)\zeta 
+(1-2\zeta+\xi)\zeta\sigma_x^2\sigma_y^2\,, \nonumber \\
\\
0 & = & \left(\ln \frac{\xi}{\zeta - \xi}-y_0\right)\frac{\zeta}{\sigma_y^2}
+ \zeta -2\xi\,.
\end{eqnarray}

With $\sigma_x$ and $\sigma_y$ fixed, $\zeta_{\rm max}$ and $\xi_{\rm
  max}$ can be used alternatively as the free parameters of
$p(\zeta,\xi)$.

\subsection{Modified Lognormal Distribution of Intrinsic 
Magnetic Field Orientations}\label{delta}

The distribution of $cos \theta_B$ follows a modified lognormal
similar to the one discussed in Appendix \ref{modlog}: let the
variable $w$ follow a Gaussian distribution with parameters $w_0,
\sigma_w$, and let $\cos\theta_B$ be related to $w$ through the
transformation
\begin{equation}
w = \ln \frac{\cos \theta_B}{1 - \cos \theta_B}\,.
\end{equation}
In this way, the PDF of $\cos \theta _B$ is 
\begin{equation}
p(\cos \theta_B) = \frac{1}{\sqrt{2\pi}\sigma}
\exp \left[-\frac{\left(\ln\frac{\cos\theta_B}{1-\cos\theta_B}-w_0\right)^2}
{2\sigma_w^2}\right] \frac{1}{\cos\theta_B(1-\cos\theta_B)}\,.
\end{equation}
The value of $\cos\theta_B$ where $p(\cos \theta_B)$ is maximized,
which we call $\cos\theta_{B \rm max}$, is obtained by requiring that
$dp/d\cos\theta_B = 0$, or, equivalently, by solving the equation
\begin{equation}
-\ln \frac{\cos\theta_B}{1-\cos\theta_B} w_0 - \sigma_w^2 +2\sigma_w^2\cos\theta_B =0\,.
\end{equation}

The standard deviation of this distribution is obtained through 
\begin{equation}
\sigma(\cos\theta_B) = \sqrt{
\int_{\cos\theta_B=0}^{1} \!\!\!\!\!\!\!\!\!\!\!\!\!\!\!\!\!\!\!
\left(cos\theta_B - \langle\cos\theta_B\rangle \right)^2 p(\cos\theta_B)d\cos\theta_B\,,
}
\end{equation}
with 
\begin{equation}
\langle \cos\theta_B \rangle = 
\int_{\cos\theta_B=0}^{1} \!\!\!\!\!\!\!\!\!\!\!\!\!\!\!\!\!\!\!
cos\theta_B p(\cos\theta_B)d\cos\theta_B\,.
\end{equation}

This is a bi-parametric distribution, with free parameters $w_0$ and
$\sigma_w$ or, equivalently, $\cos\theta_{B\rm max}$ and
$\sigma(\cos\theta_B)$.

\end{appendix}

\end{document}